\documentstyle[11pt,newpasp,twoside,epsf]{article}
\def\edcomment#1{\iffalse\marginpar{\raggedright\sl#1\/}\else\relax\fi}
\reversemarginpar

\begin{document}
\setcounter{page}{529}
\title{Uses of Linear Polarization as a Probe of Extrasolar Planet
Atmospheres} 

\author{Steven H. Saar} \affil{Harvard-Smithsonian Center
for Astrophysics, 60 Garden St., Cambridge, MA 02138} \author{Sara
Seager} \affil{Institute of Advanced Study, Einstein Dr., Princeton,
NJ 08540 and The Carnegie Institution of Washington, Dept. of
Terrestrial Magnetism, 5241 Broad Branch Rd. NW, Washington, DC 20015
}

\begin{abstract}
We point out some advantages of making observations of extrasolar
planets in linearly polarized (LP) light.  Older cool stars have quite
low levels ($\sim 10^{-4}$ to $10^{-5}$) of fractional LP, while
extrasolar planets can have relatively high fractional LP ($\sim
0.1$).  Observations in LP light can therefore significantly enhance
contrast between the planet and its parent star.  Data on LP as a
function of planetary orbital phase can be used to diagnose the
properties (e.g., composition, size, and shape) of the scatterers in
the planetary atmosphere.  We discuss the feasibility of LP
observations of extrasolar planets.
\end{abstract}

\section{Introduction}
An important ``next step" in the field of extrasolar planet research will be
to characterize their atmospheres.  The close-in extrasolar giant planets
(CEGPs) will be the first targets; indeed, the tentative detection of 
Na {\sc i} in HD 209458b has been reported (Charbonneau et al. 2002).
Old cool stars (like most known CEGP hosts) all have  very small fractional
linear polarization (LP; e.g., Leroy 1993).  At the same time, LP measurements 
are quite sensitive to the properties of scatterers in a planetary atmosphere, 
and the LP (as a fraction of the planet's total reflected light) can be 
significant (e.g., Seager, Whitney \& Sasselov 2000 [=SWS]). Thus, LP 
observations of exoplanets can potentially greatly reduce the star-planet
contrast and yield useful data on CEGP atmospheres.

\begin{figure}
\plotfiddle{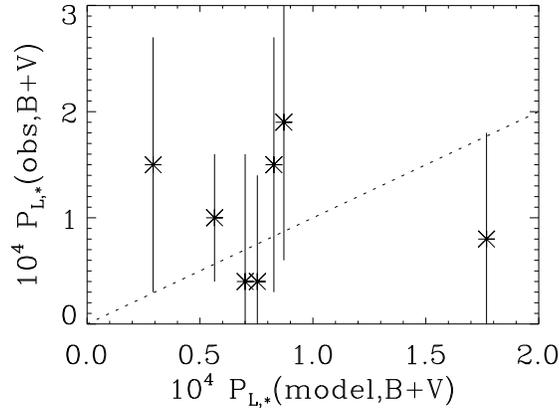}{4.5cm}{0}{100}{100}{-175}{-550} 
\caption{Broadband linear polarization $P_L$ (in the summed $B$ and $V$ bands)
from Leroy (1993)  compared with our model (based on Saar \& Huovelin 1993); 
the dashed line gives $P_L$(observed) = $P_L$(model).}
\end{figure}

\begin{figure}
\plotfiddle{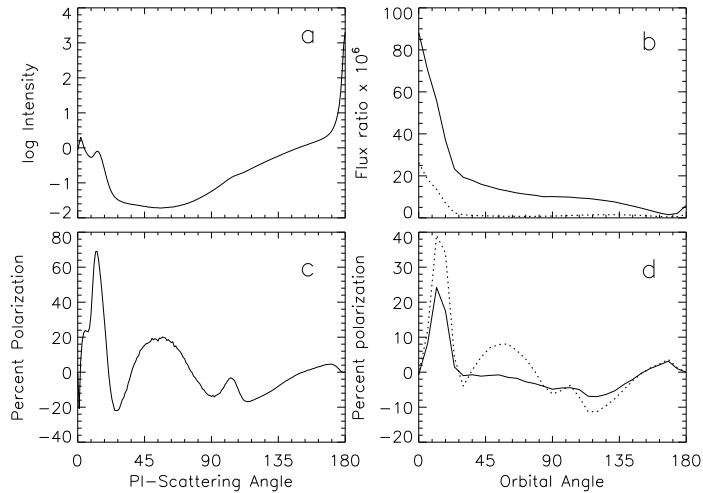}{6.0cm}{0}{50}{50}{-130}{-180}
\caption{Single scattering phase function (a) and LP probability function (c)
input into the Monte-Carlo scattering simulation (SWS).  The resulting white light phase 
function (b) and \% LP (d) are shown for the highly scattering
fiducial model (solid) and an ``absorptive" model with suppressed multiple
scattering (dashed).  Note that the LP curves (d) preserve the input particle
LP properties (c) - even when multiple scattering is high - 
much better than the light curve (b) 
preserves the scattering phase function (a).}
\end{figure}

\section{Linear Polarization Models for Stars and Exoplanets}
Since observations will be of the exoplanet-parent star system, it is important
to estimate  the LP from the host star as well.  In the absence of dust disks,
LP from cool older stars is dominated by the ``magnetic intensification" effect
from optically thick lines in a magnetic field (Leroy 1962).  This magnetic
LP has distinctive  wavelength and rotational phase properties (Huovelin \& 
Saar 1991; Saar \& Huovelin 1993).  We use the models of Saar \& Huovelin
(1993) together with active region area estimates from Saar (1996) to estimate
the maximum broadband LP (BLP) for selected CEGP host stars.  Results are
in agreement (within the large observational errors) with observed values from
Leroy (1993; Fig. 1); typically, BLP from the host star 
is on the order of 10$^{-4}$.

We model the planetary BLP as in SWS.  Our fiducial model CEGP is $a$=0.05
AU from its star,  with a radius of $R_P = 1.34 R_J$, and a gravity 
$\log g$ = 3.2.  It is covered by a uniform MgSiO$_3$ cloud, two pressure 
scale heights thick, with a log-normal particle size distribution having mean
radii of $5\ \mu$m and $\sigma = 1.5\ \mu$m. The associated scattering phase and
LP probability functions are given in Figure 2. For comparison, we also 
computed the same functions for Fe and Al$_2$O$_3$ particles (Fig. 3) 
and a Rayleigh scattering model (Fig. 4).  Figure 2
also shows results of the Monte-Carlo scattering simulation (see SWS for
details), the light, and the LP phase curves.  These results show LP
measurements can better discriminate atmospheric properties than 
the light curve for several reasons:

\begin{itemize}

\item The light curve largely reflects geometric (\% illumination) effects
(Fig. 2b), while for fractional LP (the ratio of polarized to white light),
these effects cancel out (Fig. 2d).

\item Multiple scattering smooths the light curve, blurring features
distinctive to particular particles (Fig. 2b).  The LP curve, which is 
dominated by single scattering, is much less affected in this regard (Fig. 
2d).

\item LP measurements are inherently more sensitive to the composition
(e.g., Fig. 3), size, and shape of particles than the light curve.

\end{itemize}

\begin{figure}
\plotfiddle{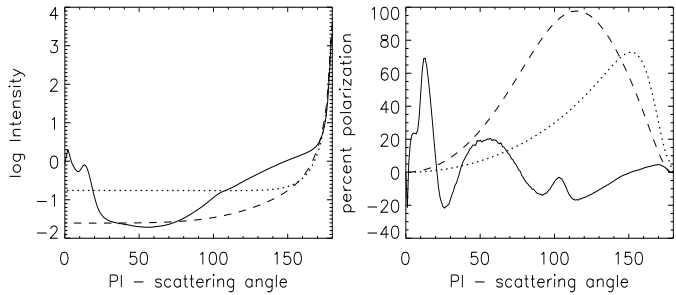}{3.5cm}{0}{50}{55}{-165}{-280}
\caption{Single scattering phase function (left) and LP probability function
(right) for three compositions: MgSiO$_3$ (solid), 
Fe (dotted) and Al$_2$O$_3$ (dashed).  LP is more sensitive to composition:
most of the distinctive features in the scattering phase function (left)
are at quite low relative intensities.}
\end{figure}

\begin{figure}
\plotfiddle{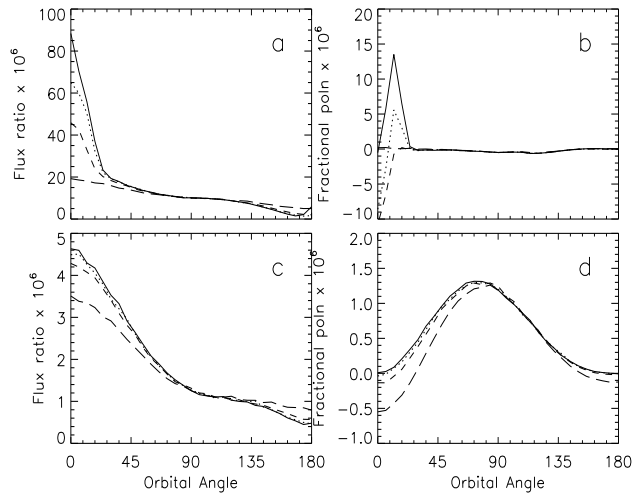}{5.70cm}{0}{45}{50}{-130}{-180}
\caption{Light curves and LP curves (as a fraction of {\it total} system 
light) for the fiducial model (a and b) and a Rayleigh scattering model
(c and d).  The Rayleigh scattering amplitude is reduced here due to 
K {\sc i} absorption; in the blue it is $\sim$4 times larger.  Once again,
model differences are clearer in LP than in white light.}
\end{figure}

\section{Feasibility and Discussion}

LP observations of CEGPs will be quite difficult; our fiducial model
suggests LP of $\sim\ $20 - 40\% of the reflected CEGP light at best (Fig. 2d).
But considerable diagnostic information is gained for this extra detection
difficulty, and the contrast between star and planet is improved
(in the $V$ band) from $I$(CEGP)/$I$(star) $\sim 5\times 10^{-5}$ (Fig. 2b)
to $P_L$(CEGP)/$P_L$(star) $\sim$ (few$\times 10^{-6}$)/(few $\times 10^{-4}) 
\sim 10^{-2}$, i.e., a factor of $\sim\ 200$. The contrast improvement, and
differential nature of the LP measurement, should aid detection 
at these low flux levels. 

Detection could be carried out by BLP polarimetry: searching for a 
weak (few \% of the total) $P_L$ signal phased to the planet's orbital period.
This method would work best in systems which are not tidally locked (e.g.,
$\upsilon$ And).  Another method would be to obtain LP spectra and use some 
multi-line method to detect the reflected, Doppler shifted spectrum
(e.g., Collier-Cameron et al. 1999).  The two
spectra will be quite distinctive: the star will display an
LP spectrum with lines characteristic of those in strong magnetic fields
(i.e., Stokes Q and U profiles), while the planet will display
a reflected (and linearly polarized in the process) spectrum of the star, 
and thus appear as a copy of the {\it unpolarized} stellar spectrum (Stokes I).
The striking difference between the two superposed spectra should
aid the identification of the exoplanet's signature.
Sensitive polarimeters and LP-capable spectrographs on large
telescopes may make such observations feasible in the near future. 

\acknowledgments This work was funded by NASA grant NAG5-10630 (SHS)
and the W.M. Keck foundation (SS).  We are very grateful to Barbara
Whitney for useful discussions and contributions.



\end{document}